\documentclass[aps,prl,reprint,groupedaddress]{revtex4-1}

\usepackage{graphicx}
\usepackage{dcolumn}
\usepackage[colorlinks=true]{hyperref}
\usepackage[caption=false,font=footnotesize]{subfig}

\usepackage{amsmath}

\usepackage{color}
\usepackage{ulem}

\begin{document}

\title{  
  Mobility
and Congestion in Dynamical Multilayer
  Networks with Finite Storage Capacity}

\author{S. Manfredi}
\email{sabato.manfredi@unina.it}
\author{E. Di Tucci}
\email{edmondo.ditucci@unina.it}
\affiliation{Department of Electrical Engineering and Information Technology, University of Naples Federico II, Naples, Italy}
\author{V. Latora}
\email{v.latora@qmul.ac.uk}
\affiliation{School of Mathematical Sciences, Queen Mary University of London, 
London E1 4NS, United Kingdom}  
\affiliation{Dipartimento di Fisica ed Astronomia, Universit\`a di Catania and INFN, I-95123 Catania, Italy}

\date{\today}
\begin{abstract}
  Multilayer networks describe well many real interconnected communication and 
  transportation systems, ranging from computer networks to multimodal
  mobility infrastructures. Here, we introduce a model in which the nodes have a limited capacity of
  storing and processing the agents moving over a multilayer network,
  and their congestions trigger temporary faults which, in turn,
  dynamically affect the routing of agents seeking for uncongested
  paths. The study of the network performance under different
  layer velocities and node maximum capacities, reveals
  the existence of delicate trade-offs between the number of served
  agents and their time to travel to destination. We provide
  analytical estimates of the optimal buffer size at which the travel
  time is minimum and of its dependence on the velocity and number of
  links at the different layers. Phenomena reminiscent of the Slower Is Faster (SIF) effect and of the Braess'
  paradox are observed in our dynamical multilayer set-up. 
  
\end{abstract}

\maketitle

Extensive studies on congestion phenomena in complex networks 
\cite{latora_nicosia_russo_2017} have highlightened the role of 
routing strategies 
\cite{Echenique2004,Echenique2005,Scellato2010,wang2014empirical,lima2016understanding,PhysRevE.73.046108} and network topologies  
\cite{de2014navigability,duch2007effect,Crucitti2004,de2014navigability,duch2007effect} on the propagation of faults and on the emergence of congestion 
\cite{guimera2002optimal,ccolak2016understanding,de2015personalized,PhysRevE.79.015101,PhysRevE.79.026112}.
Recently, multilayer network models have been introduced to better
describe complex systems composed of
interconnected networks 
\cite{kivela2014multilayer,de2013mathematical,Morris,manfredi2016}.
Diffusion processes\cite{PhysRevLett.110.028701,PhysRevLett.111.128701,PhysRevLett.118.138302} 
and congestion have also been explored in the context
of multilayer networks \cite{ren2014predicting,chodrow2016demand}. 
For instance, Ref.~\cite{Arenas2016} has shown that multiplexity 
can induce congestions that 
otherwise would not appear if the individual layers were not
interconnected. Also the effects on congestion of the mechanism
of changing layer as a function of the  velocities of
the links of two layers has been analyzed \cite{Morris}. 
However, essential aspects of the dynamics of congestions, which can play even a more important
role in multilayer systems, are still missing. 
Firstly, nodes and links have been treated as 
static entities, in the sense that the dynamical effects of the
congestion on the node queue length and agents'
routing have not been explicitely considered \cite{Morris,strano2015multiplex,Arenas2016,chodrow2016demand}. 
Secondly, the capacity of storing agents or packets at nodes has usually
been assumed infinite \cite{Arenas2016}, neglecting the dynamical
evolution of a node to be available or unavaible during time according
to whether its queue is uncongested or congested, respectively.

In this Letter, we introduce a multilayer mobility
model capturing 
both the dynamic nature of the queues at the network
nodes, and also the consequent congestion phenomena induced by the
limited storage capacity, i.e. finite {\it buffer size}, of the nodes of any
real multilayer system.  
Our model well describes the mechanisms of 
congestion occurring at the routers of a multilayer 
communication network.
The main ingredient of the model is that it 
takes into account that agents 
seek for uncongested paths during their
navigation. This makes the
routing strategy of the agents explicitly dependent on the
dynamic effects of the congestion at the nodes rather than at the links of the network
\cite{Echenique2004,Echenique2005,Scellato2010}.  
Moreover, in our model a node is unavailable
when congested and again available when becomes uncongested. Therefore, 
the onset of congestion triggers a sort of {\it temporary fault} of
the node and affects the routing of agents, which seek for uncongested
paths. In this respect, the model allows to investigate the 
effect of temporary faults on the performance of a multilayer network,
extending the analysis originally performed on single-layer networks, and limited to
permanent faults \cite{PhysRevE.66.065102,Crucitti2004,PhysRevLett.93.098701}.
Finally, the combination of
the above dynamical effects yields a {\it coevolving multilayer
  network} due to the circular argument by which the agents' routing
depends on the congestion phenomena and the latter is in turn
influenced by the agents' behavior to seek for uncongested paths.  
This leads to novel multilayer phenomena characterised by
  the existence of an optimal value of the node buffer size where the
  travel time is minimum, and which critically depends on the topologies and
  on the velocities of the different layers.

{\bf The model.~} 
The backbone of our system is modeled as a weighted multilayer
network with $L$ layers and $N^{\alpha}$ nodes at layer $\alpha$, with
$\alpha=1,2,\ldots, L$. 
Agents moving on the network can for instance represent 
information packets in a communication system: 
they are generated at the nodes
of the network and move from a node to one of its neighbours until they arrive at
their assigned final destination, where
they are removed from the network.
 The dynamics of the nodes and links is illustrated in Fig.1 of Supp.~Mat.
The main quantity of interest to characterize the state of node $i$ at layer
$\alpha$ and at time $t$, is its {\it queue length} $q_i^\alpha(t)$
which represents the number of agents at the node.  Additionally, we
assume that each node $i$ at layer $\alpha$ is characterized by its {\it maximum resource} $B_i^\alpha$,  
representing the node capability to
store agents, for instance the buffer size
of a router in a computer network. 
This implies that $q_i^\alpha(t) \leq B_i^\alpha$ for all $t$. We also
adopt First-In-First-Out (FIFO) queues, which means that agents are
processed by the queue at any given node in order of their arrival.
The basic assumption of our transportation model is that the agents
have a global knowledge of the network \cite{Scellato2010}, and
move from their origins to their destinations by following
\textit{minimum-weight} paths.  
Since our aim is to model the dynamics of agents that try to
minimise distances but also to avoid congested nodes
\cite{Echenique2004,Echenique2005,Scellato2010}, 
we assume that the link weights depend on the node queue lengths
\cite{Crucitti2004,Kinney2005}. 
Namely, the weight $w_{ij}^ {\alpha \beta}$ 
of the link connecting node $i$ at layer $\alpha$ to node $j$ at layer
$\beta$ is defined as:
\begin{equation}
  w_{ij}^ {\alpha \beta} (t)=c \cdot \gamma_{ij}^{\alpha \beta} +
  \frac{ q_{j}^\beta (t) } { B_j^\beta -  q_{j}^\beta(t) }
\label{pesi}
\end{equation}
where $\gamma_{ij}^{\alpha \beta}$ for $\alpha \neq \beta$ represents the
time of going from node $i$ at layer $\alpha$ to node $j$ at layer $\beta$,
and $c$ is the equivalent cost per unit time, so that 
$c \cdot \gamma_{ij}^{\alpha \beta}$ is the intrinsic cost of traversing the
link. Similarly, $c \cdot \gamma_{ij}^{\alpha \alpha}$ is  the intrinsic cost of
link $(i,j)$ in layer $\alpha$.
The second term in the right hand side of Eq.~(\ref{pesi}) 
represents the cost perceived by the agents and
due to the level of congestion found at the
node $j$. Notice that 
$w_{ij}^{\alpha \beta}(t) \in [ c \cdot \gamma_{ij}^{\alpha \beta}, \infty )$, with the
  weight of the link from $i$ to $j$ taking its minimum value
  $c \cdot \gamma_{ij}^{\alpha \beta}$ 
when the queue at $j$ is empty, i.e. when $q_{j}^\beta(t)=0$, while the
weight diverges when the queue is full, i.e. when $q_{j}^\beta(t)=B_j^\beta$.
At each time step, each agent computes its next move 
on the network based on the set of weights associated to the links at time $t$.  
Let us denote as $R_i^\alpha(t)$ the resulting net flow at node $i$ of
layer $\alpha$:  
\begin{equation}
 R_i^\alpha(t)=I_i^\alpha(t) +\delta I_i^\alpha(t) - O_i^\alpha(t) -\delta O_i^\alpha(t),
 \label{flussi}
\end{equation}
where $I_i^\alpha(t)$ and $O_i^\alpha(t)$ are respectively the queue
incoming and outgoing rate from and to other nodes, while 
$\delta I_i^\alpha(t)$ and $\delta O_i^\alpha(t)$ are the number of
agents generated or removed at node $i$ and layer $\alpha$, at each time.   
In particular, the output rate $O_i^\alpha(t)$ of node $i$ is related to the
capacity of the node to serve agents and route them towards other
nodes, and we assume that each node $i$ is characterized by a
maximum service rate $\hat{O}_i^\alpha$,
such that $O_i^\alpha(t)\leq \hat{O}_i^\alpha$.
Once all the values of $R_i^\alpha(t)$ are calculated, we adopt 
the following update rule of node queues:
\begin{equation}
q_i^\alpha(t+1) =
  \begin{cases}
    0 &  q_i^\alpha(t) +R_i^\alpha(t) \leq 0\\
    q_i^\alpha(t) +R_i(t)^\alpha & 0< q_i^\alpha(t)+R_i^\alpha(t)<B_i^\alpha\\
    B_i^\alpha & B_i^\alpha \leq q_i^\alpha(t) +R_i^\alpha(t)
  \end{cases}
\label{model_layer_b}
\end{equation}
for each $\alpha=1,\ldots, L$ and $i = 1, \ldots, N^{\alpha}$.  
Notice that the waiting time spent by an agent at node $i$ at
time $t$ is ${q_{i}^\alpha(t)}/{\hat{O}_i^\alpha}$, and can take a
maximum value of ${B_i^\alpha}/{\hat{O}_i^\alpha}$. 
The model takes into account that agents seek for uncongested
paths during their navigation. Moreover, agent loss may also occur 
in the model because of the congestion 
(i.e. packet loss in communication networks).
Specifically, an agent is lost at a node when it cannot be forwarded to one
of the neighbouring nodes because they are all
congested.
%
%
Summing up, the control parameters of our network model are:
$B_i^\alpha$, $\hat{O}_i^\alpha$, $\gamma_{ij}^{\alpha \beta}$,
$\forall i,j,\alpha$ and $\beta$.
%

%
\bigskip {\bf Results.~} To illustrate the rich dynamical behaviour of
the model under different control parameters,  
we consider a network with two layers ($L=2$) of
$N^{1}=150$ and $N^2=30$ nodes respectively.   
The two layers are generated as geometric random graphs by randomly
placing the nodes on the unit square and
connecting two nodes $i$ and $j$ if their Euclidean distance is
$R_{ij}<0.11$ (layer 1) or $0.11 \leq R_{ij} \leq 0.19$ (layer 2)
\cite{barthelemy2011spatial,Arenas2016}. The aim is to represent with
layer 1 a denser network with high clustering coefficient and
short-range connections, and with
layer $2$ a network of fewer nodes with long-range connections. We also assume that layer $2$
is a faster transportation system. Hence, we fix $\gamma^{1
  1}_{ij}=\gamma^{1 1}=1$ and $0<\gamma^{2 2}_{ij}=\gamma^{2 2}\leq
\gamma^{1 1}$ $\forall i,j$, so that we can explore different values
of the time ratio $\gamma=\gamma^{2 2}/\gamma ^{1 1}$ in the range
$(0, 1]$.
As the time to traverse a link is inversely proportional to
the velocity or bandwidth of the link, in the following we will refer
to $\gamma$ as the velocity ratio.  We also fix $c=1$, 
$\gamma^{1 2}_{ij}=\gamma^{1 2}=1$, $\hat{O}_i^1=\hat{O}_i^2=\hat{O}=640$ and 
$B_i^1=B_i^2=B ~\forall i,j$. 
At each time step $t$, and for each node $i$ and layer $\alpha$,
we generate with a probability $\rho$ an agent
to be delivered to each of the remaining $N^{1}+N^{2}-1$ 
of the multilayer network, such that
$\delta I_i^\alpha(t)$ is on average
equal to $\rho ( N^{1}+N^{2}-1) ~ \forall i,\alpha$.  
%

To evaluate the performance of the system we have
looked at quantities such as the average travel time and 
the number of lost agents 
for various velocity ratios $\gamma \in (0,1]$ and buffer sizes $B$. 
All the performance indexes are obtained as averages when the
network has reached a steady state condition.
\begin{figure} 
	\centering
	\includegraphics[width=0.99\columnwidth]{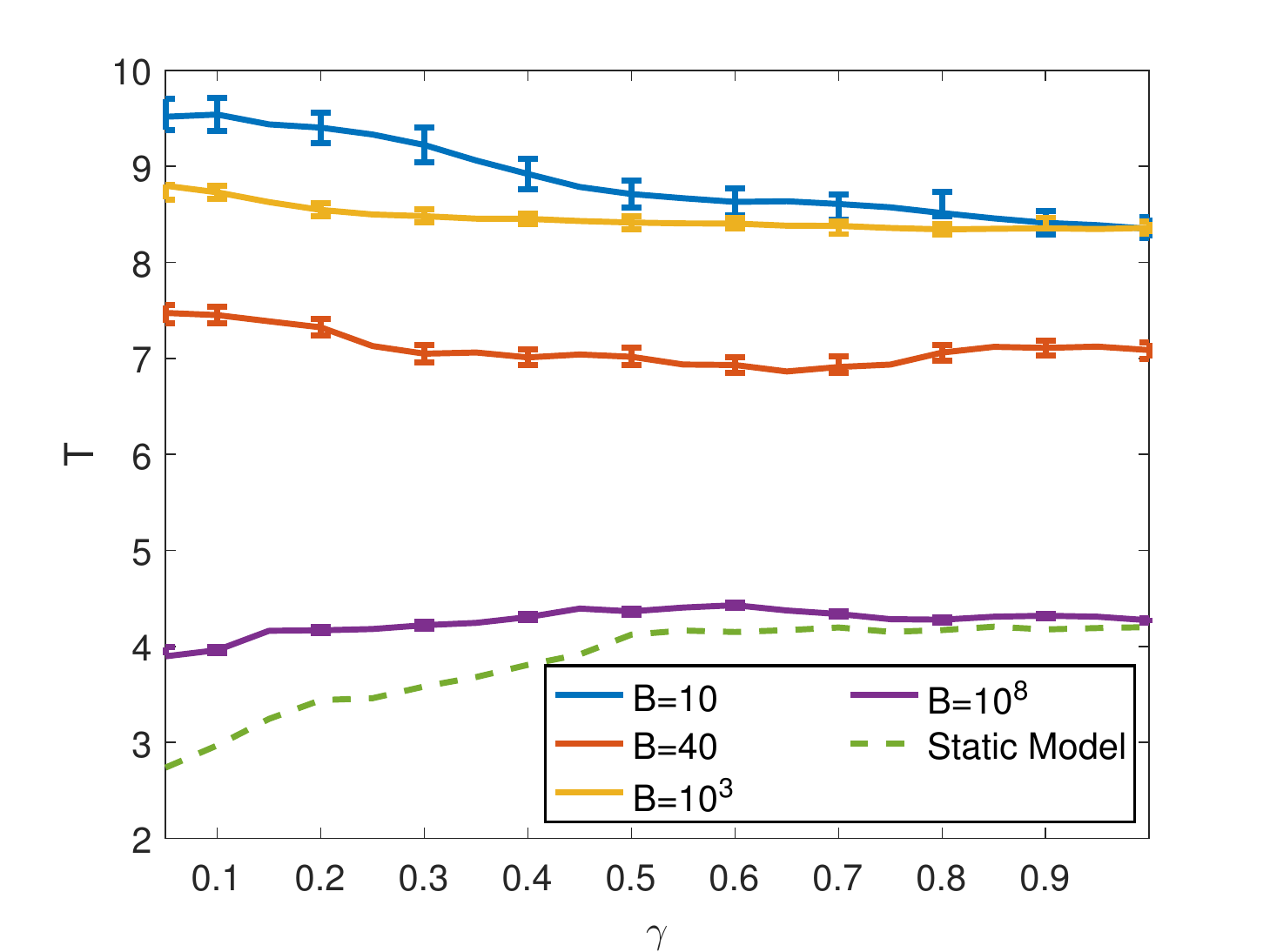} 
	\caption{Average travel time $T$ of delivered agents as a function of the velocity ratio $\gamma$ in our dynamic multilayer mobility
		model with three different values of node buffer size $B= 10, 40, 10^{3}, 10^{8}$ (continuous lines) and in the corresponding static model (dashed line). Numerical results are reported as
          symbols, with the error bars representing fluctuations
          over random agent generations and 100 different network realizations.
	\label{figs:RGG:gamma:meantime}}
\end{figure}
Fig.~\ref{figs:RGG:gamma:meantime} reports the {\it average travel
  time} $T$, defined as $T= 1 / N_a \sum_{a=1}^{N_{a}}(t^{\rm OUT}_a -
t^{\rm IN}_a)$, where $t^{\rm OUT}_a$ and $t^{\rm IN}_a$ are respectively the
times at which agent $a$ enters the network and arrives at its
destination, and $N_{a}$ is the total number of delivered agents.

The plots of $T$ as a function of $\gamma$ for buffer sizes $B= 10,
40$ and $10^{3}$ show that the travel time decreases when $\gamma$
increases, i.e. when we decrease the velocity of the faster layer $2$,
keeping fixed the velocity of layer $1$. This is an example, in a multilayer set-up,  
of the {\it Slower Is Faster} (SIF) effect reported for
several complex systems in the literature and in which the decrease of the velocity  
can yield to an improvement of the system performance \cite{SIF}.
The increase of $T$ we observe here when $\gamma$ decreases is 
due to the increasing waiting times $q(t)/ \hat{O}$ experimented by
agents along the preferential and bottlenecked paths induced by the increasingly
different velocities of the two layers. Interestingly, the effect does not occur in a {\it static model} not
accounting for the queue dynamics \cite{Morris}.  The results of the
static model, reported for comparison as dashed line, show indeed that
$T$ decreases for decreasing values of $\gamma$. As expected, our
model tends to the dashed line of the static model for very large
values of $B$ (see the curve for $B=10^8$), since the link weights in
Eq.~\eqref{pesi} become independent from the queue $q(t)$ when $B \to
\infty$.  

An analogous of the {\it Braess's paradox}, in which the addition of  
resources in terms of links leads to a worsening of network  
performance, shows up in our model in the dependence of $T$
on the buffer (i.e the addition of physical space to the nodes).
In fact, when we increase $B$ from 10 to 40 we reduce
congestion and agent losses and, as expected, we observe a drop of the
travel time (for any value of $\gamma$). However, a
  further increase of the node buffer size does not lead to an additional
  improvement of the system, as we observe, for instance at at $B=10^3$,
  an increase of the travel time $T$.
\begin{figure} 
	\centering
	\includegraphics[width=0.99\columnwidth]{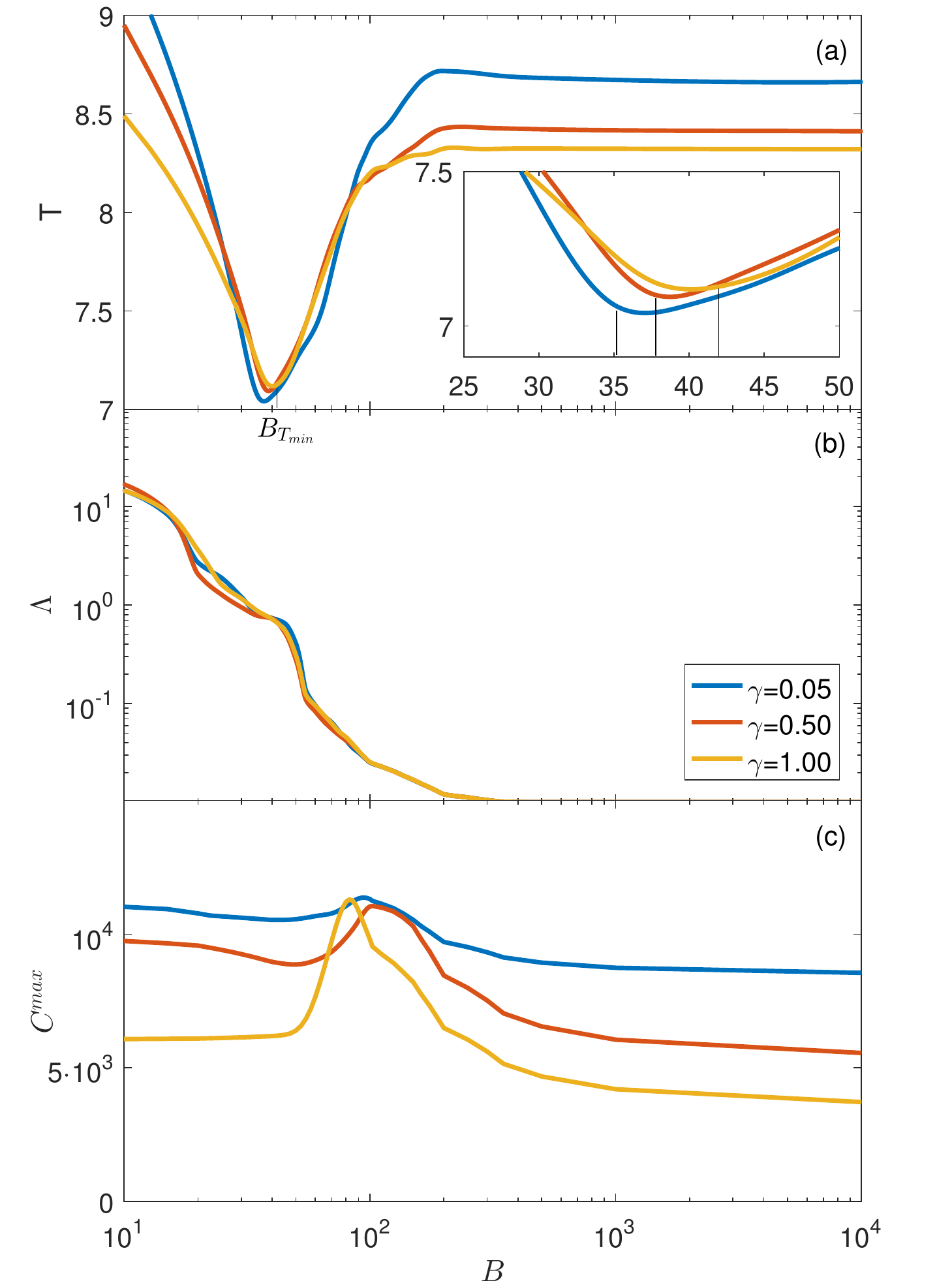} 
	\caption{Average travel time $T$ (a), percentage of agent
          loss $\Lambda$ (b), and (c) maximum betweenness
            centrality $C^{\rm max}$ for the  dynamic multilayer mobility model
                    as functions of the node buffer size $B$, and for three different
          values of velocity ratio $\gamma=0.05, 0.5, 1$.}
        \label{figs:RGG:buffer}
\end{figure}
To better highlight the non-monotonous dependence of the average travel
time from the node buffer size $B$ observed in
Fig.~\ref{figs:RGG:gamma:meantime}, in panel (a) and (b) of
Fig.~\ref{figs:RGG:buffer} we report $T$ as a function of $B$, 
together with the {\it agent
loss} $\Lambda$, defined as the percentage of agents that are unable to
reach their destination with respect to the total number of generated
agents. 
We adopt a logarithmic scale for $B$, and we show the 
results for three different velocity ratios, namely $\gamma=0.05, 0.5$ and 1.
We first describe the case $\gamma=1$. 
In the limit $B \to \infty$ the network is uncongested and $\Lambda \to 0$. 
By decreasing the value of the buffer size,  
the agent loss $\Lambda$ starts to increase at $B \simeq 10^2$
and triggers nodes alternatively being
full and empty. In this regime, the average queue length $\bar{q}$ is
well approximated by $\bar{q} \simeq B/2$ with an average waiting time
${{\bar{q}}}/{\hat{O}}\simeq B/2\hat{O}$, so that the average travel
time $T$ linearly decreases with the buffer size $B$.
This behavior holds until a value $B_{T_{\rm min}} \simeq 40$, at which the
average travel time $T$ assumes its optimal value. Finally, when the
buffer size is reduced to values $B<B_{T_{\rm min}}$, a congestion
collapse occurs with a sharp increase of agent loss.
In such high-congestion regime, the average travel time $T$
increases because of the high number of alternative but longer paths
available. 
We can notice from panel (c) of Fig.~\ref{figs:RGG:buffer}
  that, for $B\in [40,80]$, the increase in the
  travel time is related to the increase of the network maximum
  betweenness centrality $C^{\rm max}$, which is upper bounded by the node buffer
  size. This is in agreement with the results found in
  Refs.~\cite{Danila2006,Danila2007} for single-layer networks. When 
  $B>80$, the increase in $T$ is mainly due to the increased waiting time
  ${q(t)} / {\hat{O}}$ of the agents in a queue, combined with the
  nonlinear dynamics of link weights. Conversely, the maximum
  betweenness centrality $C^{\rm max}$ decreases, as the agents go through
  a lower number of more congested nodes.
The curves for $\gamma<1$ reported in Fig.~\ref{figs:RGG:buffer} 
indicate in general a worsening of the
travel time $T$ with respect to the case $\gamma=1$. Specifically, for
$B \gg B_{T_{\rm min}}$, the increase of the average travel time $T$ is
due to the increase of the waiting time $q(t)/ \hat{O}$ on 
preferential paths which gives rise to 
the Braess' paradox observed already 
in Fig. ~\ref{figs:RGG:gamma:meantime}.
For low value of buffer size $B \ll B_{T_{\rm min}}$, lower value of
$\gamma$ makes the faster layer more and more preferential and
congested such that the paths crossed by the agents include a larger
number of links and nodes of the slower layer. The effect is an
increase of $T$ for decrease of $\gamma$. In this case the number of
agent loss is almost the same but it is differently distributed along
the multilayer networks, with most occurring at the fast layer.

Differently, for buffer sizes around $B_{T_{\rm min}}$, as shown in
the inset, 
the average time $T$ is lower for higher velocity (lower $\gamma$)
as in the case of the static model. This is because 
the waiting time $q(t)/\hat{O}$ can be neglected with respect 
to the intrinsic time to traverse a link.

\begin{figure} 
	\centering
	\includegraphics[width=0.99\columnwidth]{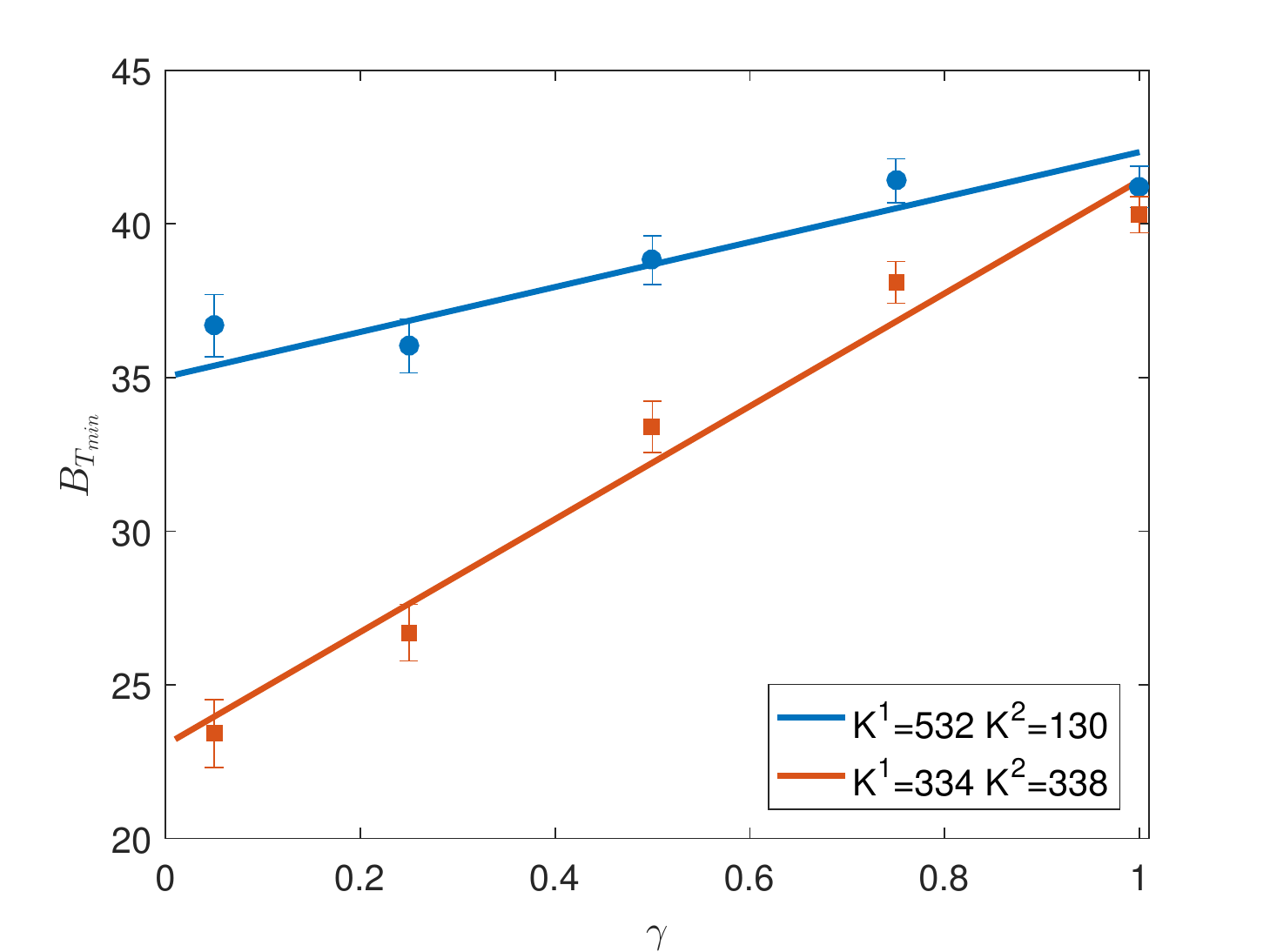} 
	\caption{Optimal buffer values $B_{T_{\rm min}}$ as a function of the
          velocity ratio $\gamma$ and for two multilayers
          (the first with $K^{1} =532$ and $K^{2} = 130$ is the one considered
          in the previous figures, while the second
          has $K^{1} = K^{2} = 334$). Numerical results are reported as
          symbols, with the error bars representing fluctuations
          over random agent generations and 100 different network realizations, while the continuous lines are the theoretical
          predictions of  Eq.~(\ref{eq:bc_estimate}).           
	\label{figs:StimaBt}}
\end{figure}

\bigskip {\bf Analytical estimations of $B_{T_{\rm min}}$.}~
The most striking result of our model 
with finite storing capacity is that the optimal value $B_{T_{\rm min}}$
of 
the buffer depends on the velocities of the layers of the
multilayer. It is therefore of outmost importance to obtain an
analytical expression of $B_{T_{\rm min}}$ as a function of $\gamma$. 
We observe that, under standard working conditions, the network
is characterized by a steady state in which the
number of generated agents equals the number of agents leaving the
network at their destination. By the Little's law \cite{little2008little} we
can then write:  
\begin{equation}
 Q=\frac{N{\bar{q}}}{\bar{\tau}}
\label{eq:little}
\end{equation}
where $Q$ is the total generated traffic per unit time, $\bar{q}$ is
the average queue length, $\bar{\tau}$ is the average time spent
by the agent over the network, and $N=N^{1}+N^{2}$. 
To evaluate $\bar{\tau}$, consider that
%
%
on average the time spent by an agent from the output of a node $i$ to
the output of its neighbour $j$ is 
$\bar{\gamma}+{{\bar{q}}}/{\hat{O}}$, namely the sum of the average
intrinsic time to cover the link $(i,j)$ and the average waiting time
at the queue ${{\bar{q}}}/{\hat{O}}$.
The value $\bar{\gamma}$ can be evaluated as
$\bar{\gamma}=\frac{\gamma^{11} K^1 + \gamma^{22} K^2 + \gamma^{12}
  K^{12}}{K^1+K^2+ K^{12}}$, where $K^{12}$, $K^1$ and $K^2$
are respectively the number of interlinks and of links in the two layers. 
Hence, the average time spent by an agent on a typical path is
$\bar{\tau} =\bar{h}(\bar{\gamma}+\frac{{\bar{q}}}{\hat{O}})$ where
$\bar{h}$ is the average number of links on the shortest path.
Since the buffer $B_{T_{\rm min}}$ is associated to the onset of the 
network congestion collapse,
i.e. when all queues are full and we have $\bar{q}/B_{T_{\rm min}}
\simeq 1$, by plugging the expression of $\bar{\tau}$ in
Eq.~\eqref{eq:little} and solving it for $\bar{q}$, we get the
following estimate for the minimum buffer size:
\begin{equation}
B_{T_{\rm min}} \simeq    \bar{q} =\frac{\bar{\gamma}Q\overline{h}}{N-\frac{Q\overline{h}}{\hat{O}}} 
	\label{eq:bc_estimate}
\end{equation}
%
%
In particular, for the case considered in our simulations, we have
$N=120$, $\overline{h}=4.1$, $Q =N(N-1)\rho$ with $\rho=0.08$,
and $\hat{O}=500$. 
For $\gamma=0.05$, $\gamma=0.5$ and $\gamma=1$ we get respectively
$B_{T_{\rm min}} \simeq 35$, $B_{T_{\rm min}} \simeq 38$ and
$B_{T_{\rm min}} \simeq 42$.  These values are reported as vertical lines in the inset of Fig.~\ref{figs:RGG:buffer} and are in good
agreement with the optimal values obtained numerically.
Eq.~(\ref{eq:bc_estimate}) highlights that a faster multilayer
network, i.e. one with lower $\bar{\gamma}$, has a lower $B_{T_{\rm
    min}}$.  This is well confirmed by Fig. \ref{figs:StimaBt}, where
we report both the numerically derived and the analytical prediction
of $B_{T_{\rm min}}$ as function of $\gamma$ for multilayer
networks with different values of $K^1$ and $K^2$.
We observe that the variation of $B_{T_{\rm min}}$ as a function of
$\gamma$ is larger when fast and slow layers have a similar number of
links. On the other hand, when the fast layer has much fewer links
than the slow layer, the value of $B_{T_{\rm min}}$ of the multilayer
network is almost independent on the difference of the
layer velocity. This also suggests that even a slight improvement in the
velocity of links of the denser layer can have better effects on
$B_{T_{\rm min}}$ than speeding up or adding new links to the
faster layer.  

\bigskip
The analytical estimate of $B_{T_{\rm min}}$ provided in this Letter 
can be useful to design efficient multilayer mobility systems. 
In particular, the model shows that an increase of network resources in
terms of the link velocities or of the buffer sizes can surprisingly lead to a worsening
of the performance.
The first behavior is reminiscent of the SIF effect \cite{SIF}
while the latter is an analogous of the Braess' paradox originally defined for 
single layer networks \cite{BraessParadox} and recently observed in 
multiplex networks as a function of the layer average degrees~\cite{Arenas2016}.  
Here we found both effects in dynamical multilayer mobility networks with
finite storing capacity.


\bigskip

\begin{acknowledgments}
This work was supported by the EPRSC-ENCORE 
Network$^{+}$ project ``Dynamics and Resilience of Multilayer Cyber-Physical Social Systems", and 
by EPSRC grant EP/N013492/1. We thank the anonymous referees for their constructive suggestions and for 
having pointed us to the SIF effect.
\end{acknowledgments}


%
\end{document}